\documentclass[useAMS,usenatbib]{mnras}
\usepackage{verbatim} 
\usepackage{natbib} 
\usepackage{amsmath} 
\usepackage{amsbsy}
\usepackage{amssymb}
\usepackage{mathrsfs} 
\usepackage{lscape} 
\usepackage{graphicx}
\usepackage{epstopdf}
\usepackage{ctable} 
\usepackage{fixltx2e} 
\usepackage{url}
\usepackage{newtxtext,newtxmath}
\usepackage[T1]{fontenc} 
\usepackage{aecompl}
\usepackage{soul}

\newcommand{\acknowledgments}{\begin{small}\section*{Acknowledgments}\end{small}}

\newcommand{\kms}{km\,s$^{-1}$}
\newcommand{\Msun}{M$_{\odot}$}

\newcommand{\Msunyr}{M$_{\odot}$yr$^{-1}$}
\newcommand{\Msunkpc}{M$_{\odot}$kpc$^{-2}$}
\newcommand{\Mbulge}{$M_{\rm BH}$--$M_{\rm bulge}$}

\newcommand{\pathL}{./}

\title[Black Holes on FIRE]
{Black Holes on FIRE: Stellar Feedback Limits Early Feeding of Galactic Nuclei}

\author[D. Angl{\'e}s-Alc{\'a}zar et al.]{
\parbox[t]{\textwidth}{\vspace{-1cm}
Daniel Angl{\'e}s-Alc{\'a}zar$^{1}$\thanks{E-mail: anglesd@northwestern.edu}, 
Claude-Andr{\'e} Faucher-Gigu{\`e}re$^{1}$,
Eliot Quataert$^{2}$,
Philip F.~Hopkins$^{3}$,
Robert Feldmann$^{4}$,
Paul Torrey$^{5}$,
Andrew Wetzel$^{3,6,7}$,
Du\v{s}an Kere\v{s}$^{8}$}
\vspace*{6pt}\\
$^1$CIERA and Department of Physics and Astronomy, Northwestern University, 2145 Sheridan Road, Evanston, IL 60208, USA\\
$^2$Department of Astronomy and Theoretical Astrophysics Center, University of California Berkeley, Berkeley, CA 94720, USA\\
$^3$TAPIR, Mailcode 350-17, California Institute of Technology, Pasadena, CA 91125, USA\\
$^4$Institute for Computational Science, University of Zurich, Zurich CH-8057, Switzerland\\
$^5$Department of Physics, MIT, 77 Massachusetts Avenue, Cambridge, MA 02139, USA\\
$^6$The Observatories of the Carnegie Institution for Science, Pasadena, CA 91101, USA\\
$^7$Department of Physics, University of California, Davis, CA 95616, USA\\
$^8$Department of Physics, CASS, University of California at San Diego, 9500 Gilman Drive, La Jolla, CA 92093, USA
\vspace{-0.4cm}
}

\date{Submitted to MNRAS, July, 2017\vspace{-0.5cm}}
\pubyear{2017}

\begin{document}
\label{firstpage}
\maketitle

\begin{abstract}
We introduce massive black holes (BHs) in the Feedback In Realistic Environments project and perform high-resolution cosmological hydrodynamic simulations of quasar-mass halos ($M_{\rm halo}(z=2) \approx 10^{12.5}$\,\Msun) down to $z=1$. 
These simulations model stellar feedback by supernovae, stellar winds, and radiation, and BH growth using a gravitational torque-based prescription tied to resolved properties of galactic nuclei. We do not include BH feedback.
We show that early BH growth occurs through short ($\lesssim$\,1\,Myr) accretion episodes that can reach or even exceed the Eddington rate. In this regime, BH growth is limited by bursty stellar feedback continuously evacuating gas from galactic nuclei, and BHs remain under-massive relative to the local \Mbulge~relation. BH growth is more efficient at later times, when the nuclear stellar potential retains a significant gas reservoir, star formation becomes less bursty, and galaxies settle into a more ordered state, with BHs rapidly converging onto the scaling relation when the host reaches $M_{\rm bulge} \sim 10^{10}$\,\Msun.
Our results are not sensitive to the details of the accretion model so long as BH growth is tied to the gas content within $\sim$100\,pc of the BH. Our simulations imply that bursty stellar feedback has strong implications for BH and AGN demographics, especially in the early Universe and for low-mass galaxies.

\end{abstract}


\vspace{-0.6cm}

\section{Introduction}

The observed connection between galaxies and central massive black holes \citep[BHs; e.g.][]{Kormendy2013,Heckman2014} poses significant challenges for galaxy formation models.
Correlations between dynamical BH mass measurements and host galaxy properties in the local universe \citep[e.g.][]{Haring2004,Savorgnan2016} have been interpreted as {\bf (1)} a non-causal consequence of hierarchical merging \citep{Peng2007,Hirschmann2010,Jahnke2011}, {\bf (2)} the causal signature of self-regulation by BH feedback \citep[e.g.][]{Silk1998,Murray2005,DiMatteo2005,Hopkins2007_BHplane}, or {\bf (3)} the result of a common gas supply for star formation and BH growth, regulated by gravitational torques \citep[e.g.][]{Kauffmann2008,Chen2013,Angles-Alcazar2013,Angles-Alcazar2015,Angles-Alcazar2017}. 
Understanding nuclear fueling in a cosmological context is a crucial step towards uncovering the nature of BH--galaxy co-evolution.

\begin{figure*}
\begin{center}
\includegraphics[width=0.41\textwidth]{\pathL/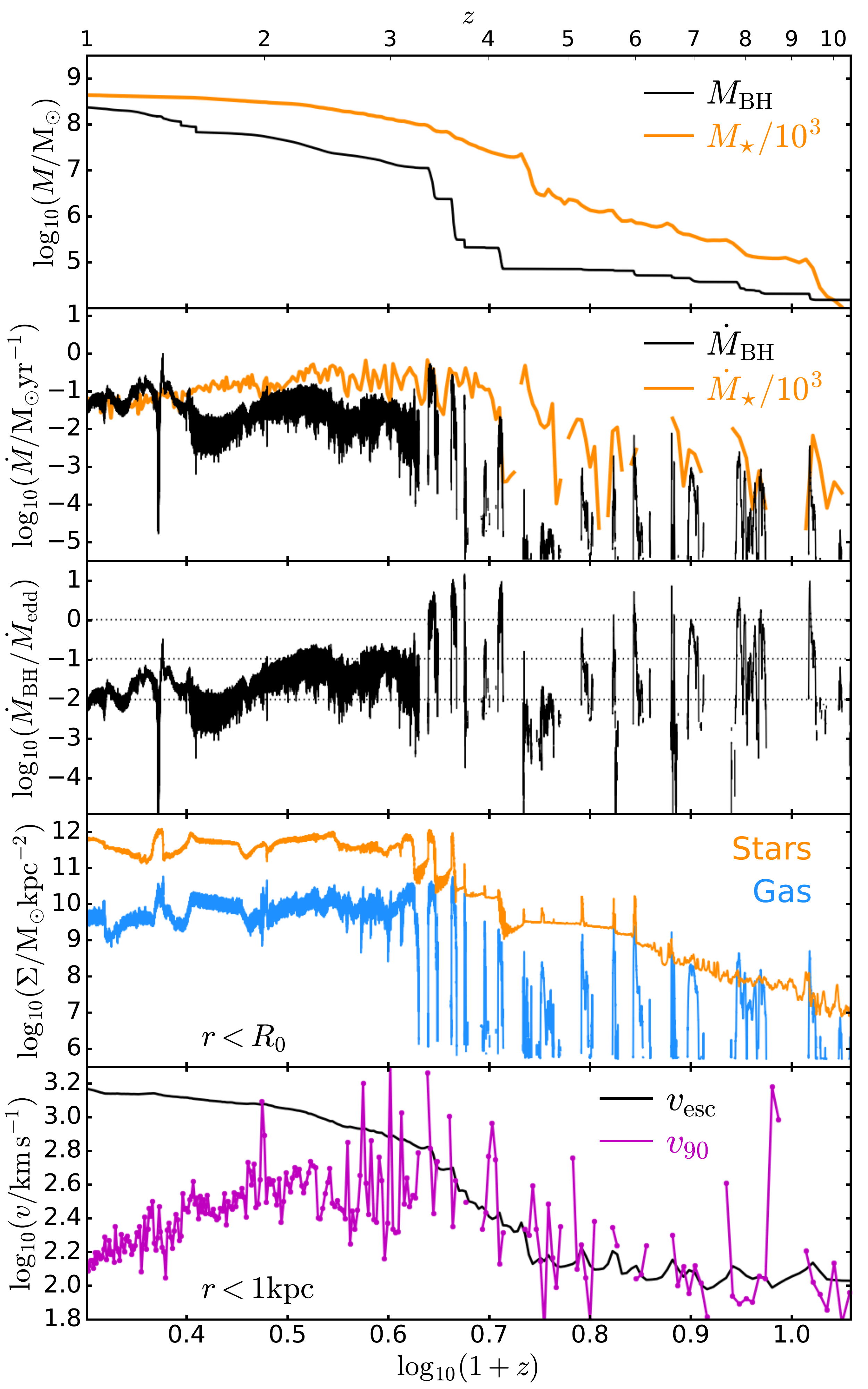}
\includegraphics[width=0.58\textwidth]{\pathL/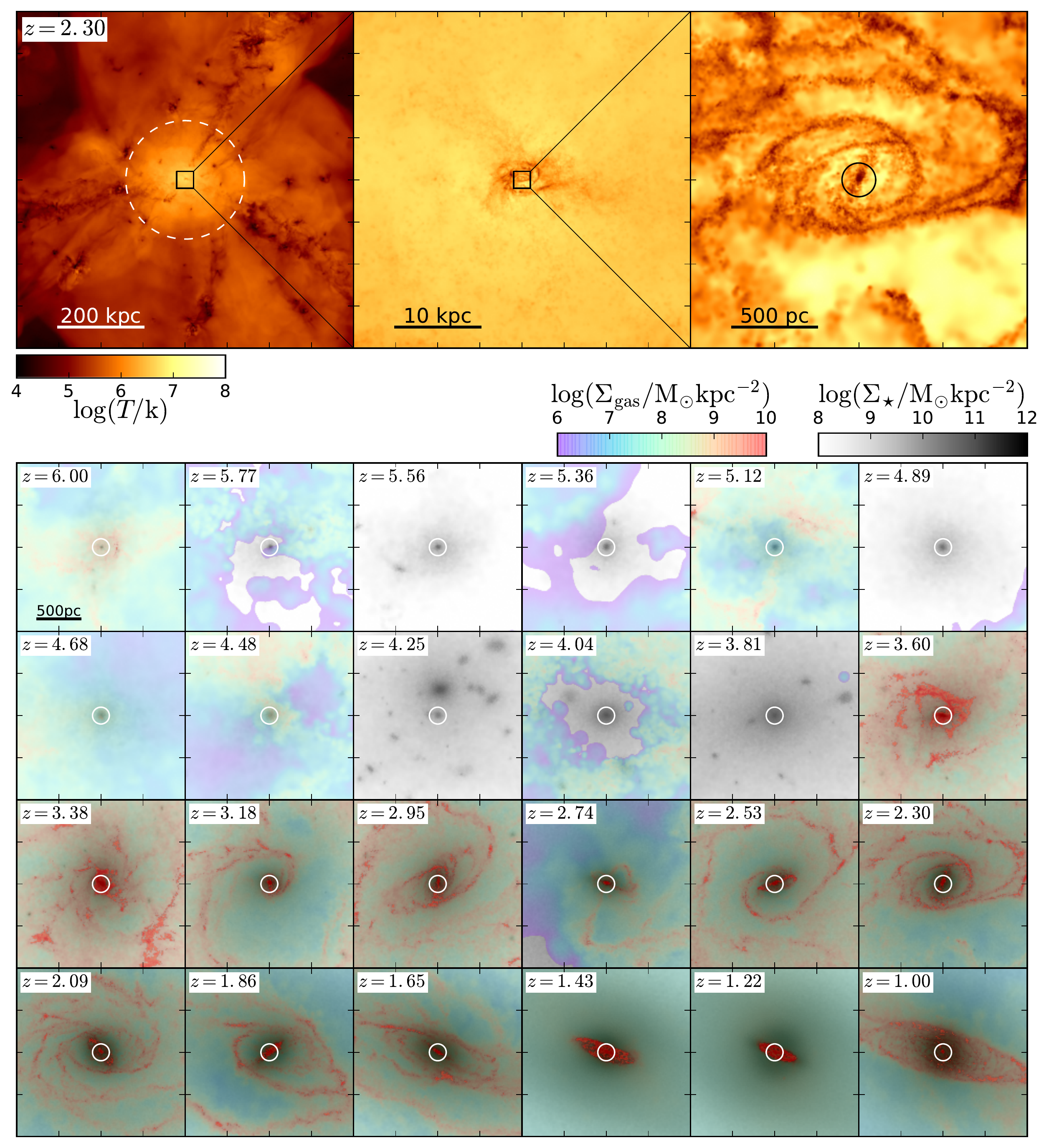}
\end{center}
\vspace{-0.3cm}
\caption{{\bf Left:} Evolution of the most massive BH in simulation {\bf A2} (a representative example).  We show, from top to bottom: {\bf (1)} BH mass, {\bf (2)} accretion rate, {\bf (3)} Eddington ratio, {\bf (4)} gas/stellar mass surface density within the variable accretion radius $R_0 \lesssim 100$\,pc, and {\bf (5)} the 90$^{\rm th}$ percentile radial velocity of outflowing gas within 1\,kpc ($v_{90}$) compared to the escape velocity at 1\,kpc ($v_{\rm esc}$).  The top two panels also indicate the total stellar mass and SFR of the host galaxy. 
{\bf Top right:} Projected mass-weighted gas temperature maps at $z=2.3$ on different scales centered on the main BH. The white dashed line indicates $R_{\rm vir}$ (left) and the black circle (right) corresponds to the central 100\,pc.  
{\bf Bottom right:} Projected gas surface density (from purple to red) overlaid on top of the stellar mass surface density (background gray scale); we show redshift evolution from $z = 6 \rightarrow 1$.  The white circles indicate the central 100\,pc. Length scales indicated on the panels are in physical units. 
At early times, $\Sigma_{\rm gas}$ fluctuates by more than three orders of magnitude owing to stellar feedback evacuating gas within the accretion radius; $\dot{M}_{\rm BH}$ can reach the Eddington rate but only intermitently during $\lesssim 1$\,Myr phases. 
More sustained BH growth begins at $z\sim 4$, when the stars dominate the gravitational potential and the nuclear gas content becomes more steady.
\vspace{-0.2cm}
}
\label{fig:bh_vs_z}
\end{figure*}

The Feedback In Realistic Environments (FIRE) cosmological ``zoom-in'' simulations\footnote{\url{http://fire.northwestern.edu}.} \citep{Hopkins2014_FIRE,Hopkins2017_FIRE2methods} offer an ideal setting to investigate the evolution of massive BHs.  By implementing stellar feedback processes on the scale of star-forming regions directly following stellar population synthesis models, the FIRE simulations reproduce a variety of galaxy
\citep{Hopkins2014_FIRE,Feldmann2017_MassiveFIRE,Ma2016_Metallicity} and CGM  \citep{Faucher-Giguere2015,Faucher-Giguere2016,Muratov2015,Hafen2017} observables.
Here, we use high-resolution cosmological hydrodynamic simulations of quasar-mass halos \citep[$M_{\rm halo} \approx 10^{12.5}$\,\Msun~at $z=2$; e.g.][]{White2012_clustering} from early times down to $z=1$ to study the impact of stellar feedback on massive BH growth.

Our simulations model the inhomogeneous, dynamic interstellar medium in the nuclear regions of galaxies ($\lesssim 100$\,pc) while self-consistently capturing mass transport from cosmological gas infall down to galactic nuclei (Fig.~\ref{fig:bh_vs_z}, top right). We rely on results from nuclear-scale simulations \citep{Hopkins2010_MultiScale,Hopkins2011_Analytic,Hopkins2016_NuclearSims} to estimate the feeding rate of the BH accretion disk based on resolved galaxy properties on scales $\lesssim$100\,pc.  
Specifically, we model the inflow rate driven by gravitational torques induced by non-axisymmetric perturbations in the stellar potential.

\vspace{-0.6cm}

\section{Simulations}\label{sec:sim}

We use the $N$-body+hydrodynamics code GIZMO\footnote{\url{http://www.tapir.caltech.edu/~phopkins/Site/GIZMO.html}} \citep{Hopkins2015_Gizmo} to re-simulate four halos from the {\bf A} series of MassiveFIRE galaxies presented in \citet{Feldmann2017_MassiveFIRE}, which did not include BH physics. 
This set of simulations covers a range of halo formation histories for halo mass $M_{\rm halo} \approx 10^{12.5}$\,\Msun~at $z=2$. Our new simulations use the updated FIRE-2 code \citep{Hopkins2017_FIRE2methods}, including the meshless finite mass (MFM) hydrodynamics solver and improvements to the accuracy of stellar feedback coupling algorithms, described therein.
We assume a standard $\Lambda$CDM cosmology consistent with observational constraints \citep[e.g.][]{PlanckCollab2015} and evolve halos down to $z=1$ with baryonic and dark matter particle masses 
$m_{\rm b} = 3.3 \times 10^4$\,\Msun~and 
$m_{\rm DM} = 1.7 \times 10^5$\,\Msun~and force softenings  
$\epsilon_{\rm gas} = 0.7$\,pc,
$\epsilon_{\rm \star} = 7$\,pc,
$\epsilon_{\rm BH} = 7$\,pc, and
$\epsilon_{\rm DM} = 57$\,pc,
where $\epsilon_{\rm gas}$ is the minimum adaptive force softening for gas (identical to the kernel smoothing scale) and $\epsilon_{\rm \star}$, $\epsilon_{\rm BH}$, and $\epsilon_{\rm DM}$ are fixed in physical units at $z<9$. 
Additionally, we use the Milky Way-mass galaxy {\bf m12i} from the FIRE-2 Latte simulation suite \citep{Wetzel2016} at three different resolution levels ($m_{\rm b} = [7,56,450] \times 10^3$\,\Msun)
for numerical convergence tests (these runs do not include BHs).  

We treat BHs as individual collisionless particles that grow through accretion and mergers \citep{Springel2005_BHmodel}. 
We model accretion as $\dot{M}_{\rm BH} = (1 - \eta) \, \dot{M}_{\rm Torque}$, where $\eta = 0.1$,  
$\dot{M}_{\rm Torque} \propto  \epsilon_{\rm T} \, f_{\rm d}^{5/2}\, M_{\rm d} \, R_{0}^{-3/2} \, M_{\rm BH}^{1/6}$ \citep[][eq. 65]{Hopkins2011_Analytic}, and $f_{\rm d}$ and $M_{\rm d}$ are the mass fraction and total mass of the disk (stars and gas) within a radial aperture $R_{0}$ enclosing 256 gas elements. 
An upper limit of 140\,pc (physical) is imposed on $R_{0}$ to avoid accreting distant gas. 
The $\epsilon_{\rm T}$ pre-factor encapsulates uncertainties in processes that affect gas transport on unresolved scales (e.g. BH feedback). We set $\epsilon_{\rm T}=2.5$ to match the observed normalization of the \Mbulge~relation at late times but $\epsilon_{\rm T}$ could in principle vary in different regimes. 
We refer to \citet{Angles-Alcazar2017} for details of the numerical implementation, including the on-the-fly bulge-disk decomposition.
BHs can exceed the Eddington rate ($\dot{M}_{\rm edd}$) by up to a factor of 10, consistent with recent simulations of super-Eddington accretion \citep[e.g.][]{Jiang2014}. Our results are largely insensitive to this limit (\S\ref{sec:post}).

We introduce one BH seed with mass $M_{\rm seed} = 1.4 \times 10^4$\,\Msun~at the location of the most bound star particle in halos with stellar mass $M_{\star}^{\rm fof} > 1000 \times M_{\rm seed}$ using a friends-of-friends algorithm \citep[e.g.][]{DiMatteo2008}. BH orbits are affected by dynamical friction, which would be underestimated owing to finite mass resolution. In order to more physically model BH dynamics, 
each BH is given an initial ``dynamical mass'' $m_{\rm BH} = 300 \times m_{\rm b} \approx 60 \times m_{\rm DM}$ independent of the physical $M_{\rm BH}$ set by accretion. 
Once $M_{\rm BH}$ reaches $m_{\rm BH}$, both remain equal for the rest of the simulation. 
BH feedback is intentionally disabled to allow for a clean exploration of the impact of stellar feedback on BH growth.
BH properties are saved at every time-step, yielding a time resolution in $\dot{M}_{\rm BH}$ of up to $10^3$\,yr.

\vspace{-0.6cm}

\section{Results}

\subsection{Representative black hole accretion history}

Fig.~\ref{fig:bh_vs_z} (left) shows the growth histories of the main galaxy in simulation {\bf A2} and its central BH.  
At early times, {\bf A2} experiences intense bursts of star formation, growing from $M_* \sim 10^{7} \rightarrow 10^{10}$\,\Msun~in the redshift range $z=10 \rightarrow 4$.  During this early period, the total star formation rate (SFR) can reach $\dot{M}_* \sim 10$--400\,\Msunyr. 
Stellar feedback drives large scale winds with 90th-percentile velocity $v_{\rm 90} \sim 10^2$--$10^3$\,\kms~(within 1\,kpc), evacuating a large fraction of ISM gas and temporarily shutting down star formation \citep{Muratov2015,Angles-Alcazar2016}. 
Feedback-driven outflows/inflows cause significant radial stellar migration seen as fluctuations in $\Sigma_{*}$ \citep{El-Badry2016}.
As the galaxy grows to $M_{*} > 10^{10}$\,\Msun, star formation becomes less bursty and rapid outflow events become less frequent \citep[e.g.][]{Ma2017_DiverseGradients}.  
Sustained star formation at rates $\dot{M}_* \sim 100$\,\Msunyr~grows the stellar mass to $M_* = 4\times 10^{11}$\,\Msun~by $z=1$ while maintaining a substantial nuclear gas reservoir ($\Sigma_{\rm gas} \sim 10^{10}$\,\Msunkpc).

Despite the rapid growth of the host galaxy, the central BH grows by less than one order of magnitude by $z=4$.  Early BH growth occurs through sporadic accretion episodes, reaching $\dot{M}_{\rm BH} \sim 0.01$\,\Msunyr~when the nuclear gas surface density peaks at $\Sigma_{\rm gas} \sim10^8$--$10^9$\,\Msunkpc.
BH accretion can reach or even exceed the Eddington rate during short phases ($\lesssim 1$\,Myr), but the total mass accumulated is limited by the availability of gas to accrete over longer timescales. At early times, $\Sigma_{\rm gas}$ fluctuates by more than three orders of magnitude owing to bursty stellar feedback evacuating gas from the nucleus for $\sim 10$--100\,Myr intervals \citep[e.g.][]{Torrey2017}. Efficient BH growth requires a sustained nuclear gas reservoir, only achieved at $z<4$ when the stellar component dominates the gravitational potential and gas is more effectively retained in the nucleus (in other simulated halos, efficient BH growth can be delayed until $z \sim 2$).  
Thumbnails in Fig.~\ref{fig:bh_vs_z} (bottom right) show the evolution of the gas and stellar components in the central 2\,kpc from $z = 6 \rightarrow 1$, illustrating the transition from {\bf (1)} the irregular morphology, bursty star formation, and highly dynamic conditions prevalent at early times to {\bf (2)} the well-defined stellar potential, more steady star formation, and long-lived nuclear gas disk enabling sustained BH growth at later times.

We find $\dot{M}_{\rm BH}\sim 0.1$\,\Msunyr~frequently at $z<4$, corresponding to a bolometric luminosity $L_{\rm bol}\sim 6 \times 10^{44}$\,erg\,s$^{-1}$ (for a 10\% radiative efficiency), while rare accretion episodes can reach $\dot{M}_{\rm BH}\sim 1$\,\Msunyr.
Our simulations thus predict that moderate luminosity active galactic nuclei (AGN) at high redshift can be stochastically fueled and do not require major mergers events, in agreement with observations \citep[e.g.][]{Kocevski2012}. 
The gas accreted by BHs during the late-time, more steady growth phase is heavily metal-enriched and a large fraction ($\gtrsim 50$\%) has been processed in earlier generations of stars, but not necessarily in the nucleus.

\begin{figure}
\begin{center}
\includegraphics[width=0.43\textwidth]{\pathL/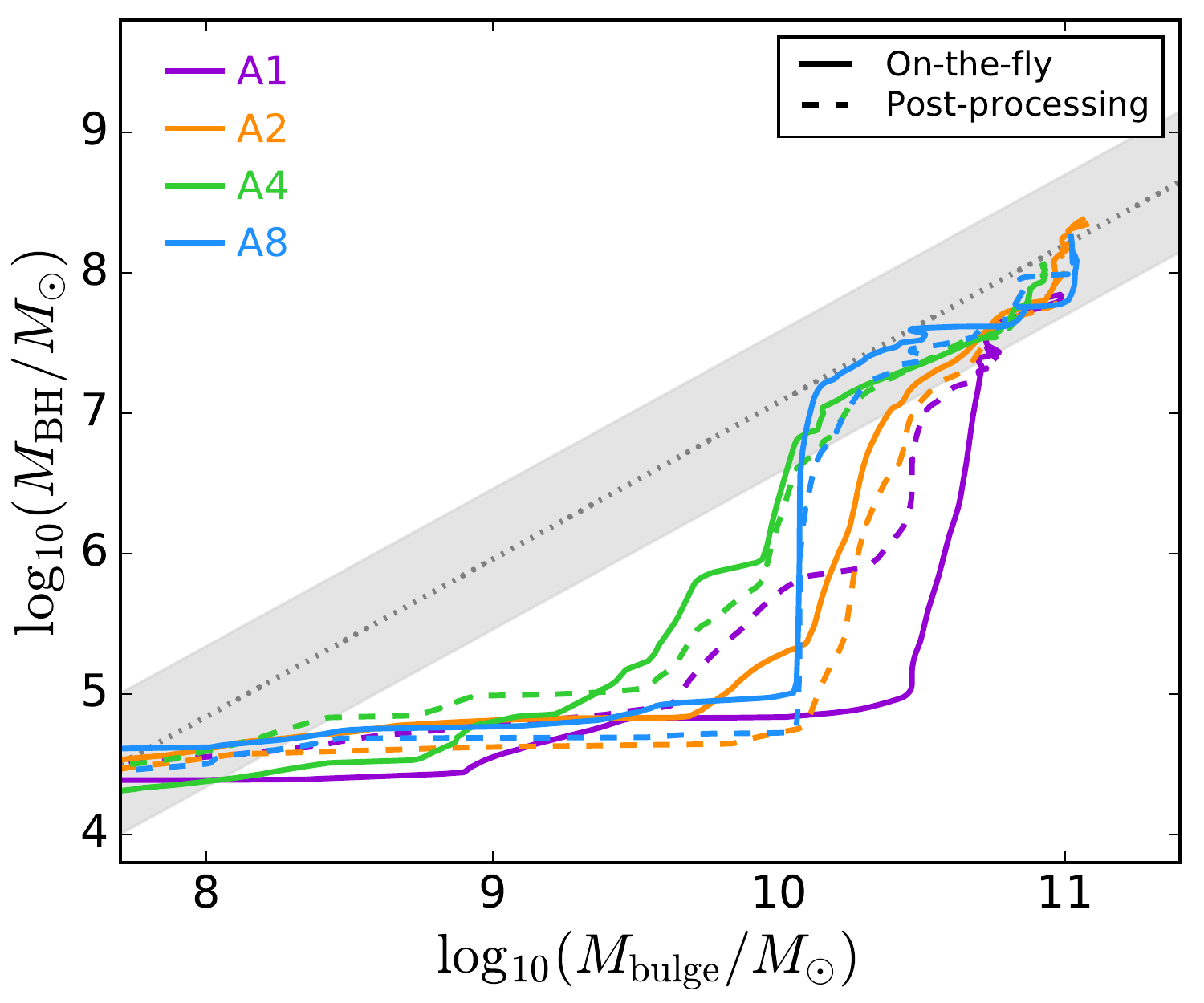}
\end{center}
\vspace{-0.3cm}
\caption{Evolutionary tracks of BHs and galaxies in the \Mbulge~plane from early times down to $z=1$ for the main halo in each simulation.  
Solid lines show simulation results while dashed lines correspond to post-processing calculations (see \S \ref{sec:correlations}). The black dotted line and gray shaded area indicate the \citet{Haring2004} relation and 0.5 dex scatter in $M_{\rm BH}$.
Black holes are under-massive at early times and converge onto the scaling relation when their hosts reach $M_{\rm bulge} \sim 10^{10}$\,\Msun.
\vspace{-0.2cm}
}
\label{fig:MM}
\end{figure}

\begin{figure}
\begin{center}
\includegraphics[width=0.48\textwidth]{\pathL/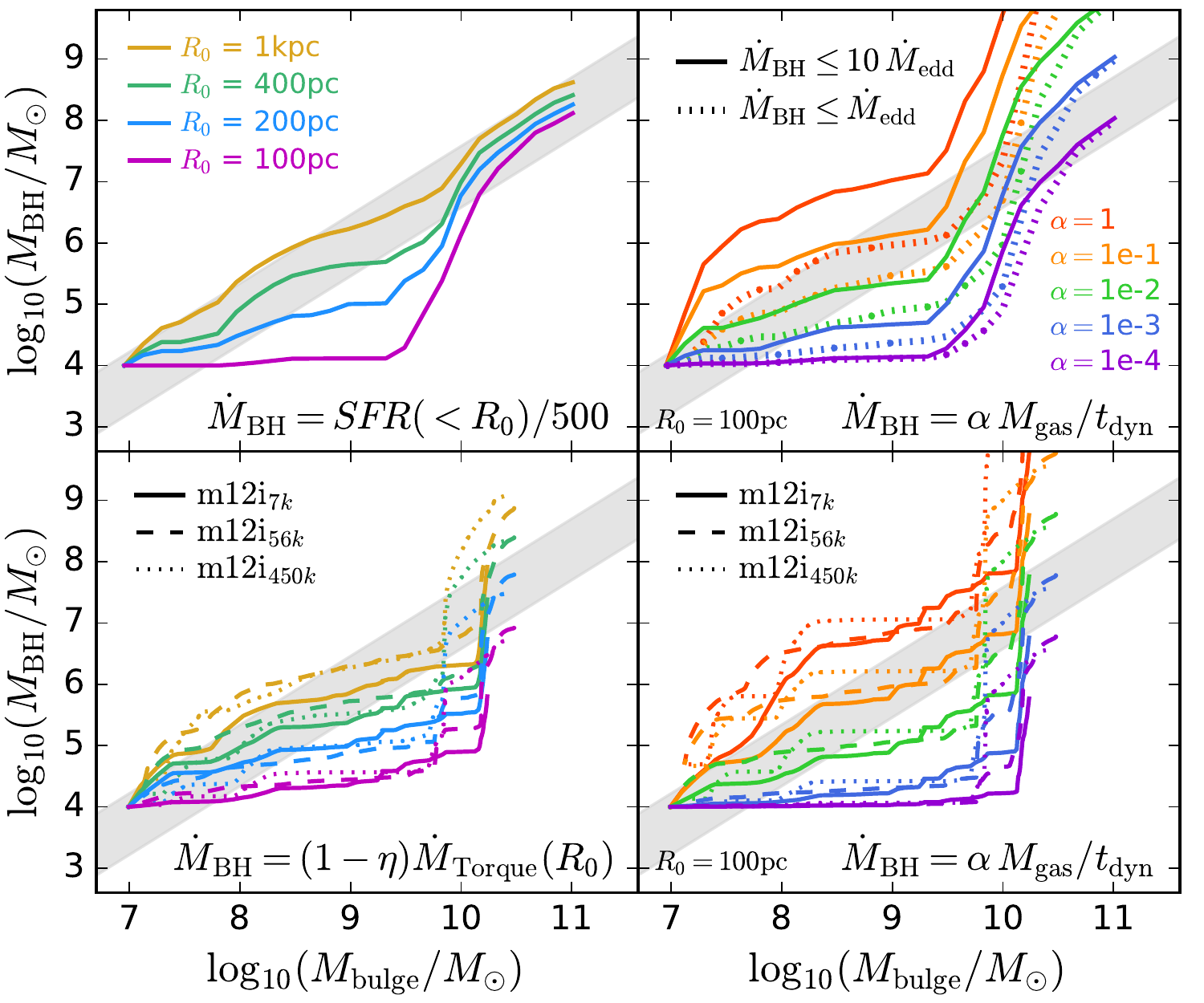}
\end{center}
\vspace{-0.3cm}
\caption{{\bf Top panels:} Evolutionary tracks in the \Mbulge~plane averaged over our four quasar-mass halos, where we compute BH growth in post-processing as
$\dot{M}_{\rm BH} = \dot{M}_{*}(R_0)/500$ for the SFR within different apertures $R_0 = 100$\,pc--1\,kpc (left), and
$\dot{M}_{\rm BH} = \alpha \, M_{\rm gas}/t_{\rm dyn}$ for different $\alpha$ values, where $t_{\rm dyn} \equiv (R_{0}^{3}/GM_{\rm tot})^{1/2}$ and $M_{\rm tot}$ is the total mass within $R_0=100$\,pc (right).
{\bf Bottom panels:} Convergence tests for galaxy {\bf m12i} at three resolution levels, where we compute 
$\dot{M}_{\rm BH} = (1 - \eta) \times \, \dot{M}_{\rm Torque}$ for different $R_0$ (left) and 
$\dot{M}_{\rm BH} = \alpha \, M_{\rm gas}/t_{\rm dyn}$ for $R_0=100$\,pc and different $\alpha$ values (right).
We limit accretion to $\dot{M}_{\rm BH} \leq 10\,\dot{M}_{\rm edd}$ in all cases, except in the top right panel where we also show $\dot{M}_{\rm BH} \leq \dot{M}_{\rm edd}$ (dotted lines).
The gray shaded area shows the \citet{Haring2004} relation. 
Stellar feedback limits early BH growth when $\dot{M}_{\rm BH}$ is tied to the gas content within $\sim 100$\,pc, 
but $\dot{M}_{\rm BH}$ could be high enough to grow early massive BHs more quickly than in our fiducial calculations if $\alpha$ is temporarily elevated.
\vspace{-0.2cm}
}
\label{fig:MM_tests}
\end{figure}

\vspace{-0.3cm}

\subsection{Black hole--host correlations}\label{sec:correlations}

Fig.~\ref{fig:MM} shows the evolution of BHs and hosts in the \Mbulge~plane from early times down to $z=1$, where $M_{\rm bulge}$ is the bulge mass within the stellar effective radius (from a 3D kinematic decomposition; e.g. \citealt{Angles-Alcazar2014}). 
As expected from Fig.~\ref{fig:bh_vs_z}, the central BH in galaxy {\bf A2} grows very little at early times, departing from the local scaling relation as the host galaxy grows.  When the stellar bulge reaches $M_{\rm bulge} \sim 10^{10}$\,\Msun, the BH quickly converges onto the scaling relation \citep[see also, e.g.][]{Dubois2015_SNa}.
Similar tracks are seen for our {\bf A1}, {\bf A4}, and {\bf A8} simulations.
The rapid increase in $M_{\rm BH}$ seen for {\bf A8} corresponds to a merger-triggered super-Eddington growth phase.

The dashed lines in Fig.~\ref{fig:MM} show evolutionary tracks from post-processing calculations based on the same simulations, assuming that the BH is always located at the (dark matter + baryons) density center of the halo computed with AHF \citep{Knollmann2009_AHF}.
Following \citet{Angles-Alcazar2015}, we integrate $\dot{M}_{\rm BH}$ using simulation snapshots available every $\sim 10$--25\,Myr, evaluating $\dot{M}_{\rm Torque}$ for a variable aperture $R_0$ as in our on-the-fly calculation (\S\ref{sec:sim}).  This analysis is insensitive to the exact definition of halo/galaxy center and approximates very well our on-the-fly results. 
Since in our simulations BHs can move away from galaxy centers owing to dynamical interactions \citep[e.g.][]{Tremmel2015_DynFriction,Biernacki2017}, we conclude that BH dynamics is not the dominant factor responsible for the early suppression of BH growth.

\vspace{-0.4cm}

\subsection{Post-processing analysis}\label{sec:post}

Fig.~\ref{fig:MM_tests} shows several post-processing variations of our BH growth analysis. 
The top left panel shows \Mbulge~evolutionary tracks down to $z=1$ averaged over our four quasar-mass halos assuming that $\dot{M}_{\rm BH}$ is proportional to the SFR within a fixed physical radius at all times, $\dot{M}_{\rm BH} = \dot{M}_{*}(R_0)/500$, where the normalization factor is the same for all $R_0$ (roughly the local observed $M_{\rm BH}/M_{\rm bulge}$ ratio).  
The characteristic transition in $M_{\rm BH}$ at $M_{\rm bulge} \sim 10^{10}$\,\Msun~is recovered when the post-processing aperture is comparable to the on-the-fly calculations ($R_0 = 100$\,pc). However, early BH growth is over-estimated when using a larger aperture ($R_0 = 1$\,kpc) linking $\dot{M}_{\rm BH}$ to the galaxy-scale gas reservoir.
The top right panel of Fig.~\ref{fig:MM_tests} shows average evolutionary tracks assuming that $\dot{M}_{\rm BH}$ is proportional to the total gas mass within $R_0=100$\,pc accreted per dynamical time, $\dot{M}_{\rm BH} = \alpha \, M_{\rm gas}/t_{\rm dyn}$, where we vary the normalization $\alpha$ and $\dot{M}_{\rm BH}$ is limited to either $10\,\dot{M}_{\rm edd}$ (solid lines) or $\dot{M}_{\rm edd}$ (dotted lines).
Predictions from this free-fall accretion estimator are in good agreement with $\dot{M}_{\rm Torque}$ for $\alpha = 10^{-4}$--$10^{-3}$. 
The $\alpha = 1$ case represents the maximum BH fueling rate (for a given Eddington limit). Even in this case, the evolutionary tracks retain the characteristic transition from slower to faster growth at $M_{\rm bulge} \sim 10^{10}$\,\Msun.

The bottom left panel of Fig.~\ref{fig:MM_tests} shows \Mbulge~evolutionary tracks down to $z=0$ for the Milky Way-mass galaxy {\bf m12i} at three different resolution levels, where we integrate $\dot{M}_{\rm BH} = (1 - \eta) \times \, \dot{M}_{\rm Torque}$ in post-processing for different $R_0$ (the {\bf m12i$_{7k}$} run reaches a mass resolution $\sim 5 \times$ finer than our quasar-mass halos).
Small accretion radii ($R_0 \lesssim 200$\,pc) produce evolutionary tracks in good agreement with Fig.~\ref{fig:MM}, while larger apertures overestimate $\dot{M}_{\rm BH}$. 
Despite a trend of earlier BH growth in lower-resolution calculations, evolutionary tracks are well converged for a wide range of $R_0$ and a factor $64\times$ difference in mass resolution. We have also evolved our four quasar-mass halos at $8\times$ lower mass resolution and confirmed that our main conclusions are not sensitive to resolution. 
The bottom right panel of Fig.~\ref{fig:MM_tests} shows the evolutionary tracks obtained for {\bf m12i} evaluating $\dot{M}_{\rm BH} = \alpha \, M_{\rm gas}/t_{\rm dyn}$ within $R_0=100$\,pc, demonstrating good numerical convergence with resolution for different $\alpha$ values. 
The similar shape and normalization relative to the average evolutionary tracks for our quasar-mass halos down to $z=1$ (top right panel) suggest that the characteristic $\dot{M}_{\rm BH}$ transition at $M_{\rm bulge} \sim 10^{10}$\,\Msun~is independent of redshift, but a more comprehensive analysis will be needed to confirm this.

\vspace{-0.5cm}
\section{Discussion and Conclusions}

We present the first cosmological simulations coupling the FIRE stellar feedback physics with a model for massive BH growth based on gravitational torques. 
By resolving the inner $\sim$100\,pc of galaxies, we show that stellar feedback regulates the gas reservoir in galactic nuclei, which can severely limit early BH growth.
Efficient BH growth begins when stars dominate the gravitational potential in the nucleus and star formation becomes less bursty, roughly when $M_{\rm bulge} \sim 10^{10}$\,\Msun. At this stage, the galaxy center becomes well defined and the escape velocity at 1\,kpc exceeds that of stellar feedback-driven winds.   
This evolution in BH fueling mode roughly coincides with galaxy-scale transitions found previously in FIRE simulations, in which early bursty star formation transitions to long-lived gaseous disks with more time-steady star formation in more massive and lower-redshift galaxies \citep{Muratov2015,Angles-Alcazar2016,Faucher-Giguere2017,Hayward2017,Ma2017_DiverseGradients,Ma2017_StellarDisk}. Dwarf galaxies experience bursty star formation down to $z=0$ \citep[e.g.][]{Angles-Alcazar2016}, suggesting that BH growth will also be inefficient in low-redshift dwarfs. In such galaxies, inefficient BH growth may correlate with the formation of dark matter cores and stellar size fluctuations driven by stellar feedback \citep[e.g.][]{Chan2015,El-Badry2016}.

Recent simulations based on Bondi-like accretion and parameterized star formation-driven kinetic winds \citep{Costa2014} or delayed cooling thermal supernova feedback  \citep{Dubois2015_SNa,Bonoli2016,Bower2017,Habouzit2017,Prieto2017} also find suppressed $\dot{M}_{\rm BH}$ in low-mass galaxies, in qualitative agreement with our results. 
We find that the early suppression of BH growth by stellar feedback is generic to models in which $\dot{M}_{\rm BH}$ is tied to the nuclear gas content, provided that the effects of stellar feedback in the nucleus are resolved. 
Simulations that model $\dot{M}_{\rm BH}$ based on the larger-scale ($\gtrsim 500$\,pc) galactic gas reservoir can greatly overestimate early BH growth.
We note, however, that Bondi accretion may inhibit the growth of low-mass BHs even in the presence of a continuous gas supply owing to the strong dependence on BH mass, $\dot{M}_{\rm Bondi} \propto M_{\rm BH}^{2}$ \citep[e.g.][]{Angles-Alcazar2015}. 
In our simulations, most of the gas in the central $\sim$100\,pc is cold and rotationally supported when BHs grow efficiently, which justifies the use of the gravitational torque model.
Nonetheless, $\dot{M}_{\rm BH}$ could be higher than predicted by our fiducial gravitational torque model at early times. For sufficiently high accretion efficiency per free-fall time ($\alpha >1$\%), there would be enough gas in early nuclei to rapidly grow BHs to the local \Mbulge~relation.

Our fiducial simulations predict under-massive BHs in low-mass galaxies, in agreement with observations of the local \Mbulge~relation \citep[e.g.][]{Graham2013,Savorgnan2016}, and substantial scatter in $M_{\rm BH}$ at $M_{\rm bulge} \sim 10^{10}$\,\Msun~owing to rapid convergence onto the scaling relation once BHs start growing efficiently. In future work, it will be interesting to more systematically investigate the implications of stellar feedback for the growth of $z>6$ quasars \cite[e.g.][]{Mortlock2011}, over-massive relic BHs \citep[e.g.][]{McConnell2011}, and $M_{\rm BH}$ measurements in active dwarfs \citep[e.g.][]{Jiang2011,Reines2013}.
The transition in BH fueling mode driven by stellar feedback also has direct implications for AGN demographics \citep[e.g.][]{Kauffmann2003,Aird2017} and for massive BH mergers and their observability by future gravitational wave missions, such as the Laser Interferometer Space Antenna \citep[LISA;][]{Amaro-Seoane2017}. BHs are expected to be under-massive in low-mass satellites and experience reduced dynamical friction. The frequency and mass scale of massive BH mergers may thus be affected by stellar feedback.

After the early phase regulated by stellar feedback, BHs evolve along the \Mbulge~relation without the need for large-scale AGN feedback self-regulation, in agreement with previous simulations with simpler subgrid ISM \citep[][]{Angles-Alcazar2013,Angles-Alcazar2015}. 
Simulations implementing gravitational torque-driven BH growth and feedback on $\sim$kpc scales indicate that large-scale AGN feedback may have a weak effect on the scaling relations, suppressing the growth of galaxies and BHs by a similar amount \citep{Angles-Alcazar2017}. 
Nonetheless, BH feedback on smaller scales may play a significant role in determining $\dot{M}_{\rm BH}$, which could affect some of our results.
Future work should address the impact of AGN feedback in cosmological simulations implementing the stellar physics and dynamic range necessary to resolve galactic nuclei.

\vspace{-0.5cm}
\acknowledgments 
{
DAA acknowledges support by a CIERA Postdoctoral Fellowship.
CAFG was supported by NSF grants AST-1412836 and AST-1517491, NASA grant NNX15AB22G, STScI grant HST-AR-14562.001, and CXO grant TM7-18007X.
EQ was supported by NASA ATP grant 12-ATP-120183, a Simons Investigator award from the Simons Foundation, and the David and Lucile Packard Foundation. 
Support for PFH was provided by an Alfred P. Sloan Research Fellowship, NASA ATP Grant NNX14AH35G, and NSF grant AST-1411920 and CAREER award \#1455342. 
RF acknowledges support from the Swiss National Science Foundation (grant no 157591).
PT acknowledges support from NASA through Hubble Fellowship grant HST-HF-51384.001-A awarded by STScI.
AW was supported by a Caltech-Carnegie Fellowship, the Moore Center for Theoretical Cosmology and Physics at Caltech, and NASA grant HST-GO-14734 from STScI.
DK was supported by NSF Grant AST1412153 and a Cottrell Scholar Award from the Research Corporation for Science Advancement. 
The simulations were run using XSEDE, supported by NSF grant ACI-1053575, and Northwestern University's compute cluster ``Quest''.
}

\vspace{0.5cm}

\bibliography{./bhletter}

\end{document}